\documentclass[%
twocolumn,
groupedaddress,
amsmath,amssymb, nofootinbib
]{revtex4-2}

\usepackage[english]{babel}

\usepackage{algorithm2e}
\usepackage{natbib}
\usepackage{color}

\usepackage{amsmath}
\usepackage{amsthm}
\usepackage{amssymb}
\usepackage{graphicx}
\usepackage{subcaption}
\usepackage[mathscr]{euscript}
\usepackage[colorlinks=true, allcolors=blue]{hyperref}

\newtheorem{theorem}{Theorem}

\begin{document}

\title{On the equivalence between nonlinear graph-based dynamics and linear dynamics on higher-order networks}
\author{Lucas Lacasa}
\affiliation{Institute for Cross-Disciplinary Physics and Complex Systems (IFISC, CSIC-UIB),\\Campus UIB, 07122 Palma de Mallorca, Spain}

\begin{abstract}
In network science, collective dynamics of complex systems are typically modelled as
(nonlinear, often including many-body) vertex-level update rules evolving over a graph interaction
structure. In recent years, frameworks that explicitly model such higher-order interactions in the interaction backbone (i.e. hypergraphs) have been advanced, somehow shifting the imputation of the effective nonlinearity from the dynamics to the interaction structure.
In this work we discuss such structural--dynamical representation duality, and investigate how and when a nonlinear dynamics defined on the vertex set of a graph
allows an equivalent representation in terms of a linear dynamics defined on the state space of a sufficiently richer, higher-order interaction structure. Using Carleman linearisation arguments, we show
that finite polynomial dynamics defined in the $|V|$ vertices of a graph admit an exact representation as linear dynamics on the state space of an hb-graph of order $|V|$, a combinatorial structure that extends hypergraphs by allowing vertex multiplicity, where the specific shape of the nonlinearity indicates whether the hb-graph is either finite or infinite (in terms of the number of hb-edges).
For more general analytic nonlinearities, exact linear representation always require an hb-graph of infinite size, and its finite-size truncation provides an approximate representation of the original nonlinear graph-based dynamics.
\end{abstract}

\maketitle

A fundamental modelling question in network dynamics asks whether simple interaction structures --i.e. graphs-- are sufficiently expressive to account for the emergent collective dynamics observed in real-world complex systems, or whether it is necessary to consider more sophisticated interaction structures, such as simplicial complexes or hypergraphs. While many recent efforts in the network science community advocate for the latter (see e.g. \cite{Battiston2020, SIAM, Battiston25} for recent reviews), recently Peixoto {\it et al} \cite{Peixoto} have argued that there is a fundamental conflation between high-order interactions and high-order networks, and that the former do not necessarily require the latter --e.g. insofar hypergraph-encoded interactions can  already be acommodated within graph-based models--. Note that similar reservations had been raised before for particular cases \cite{maxi, lambiotte}.\\
Higher-order interactions occur when three or more variables jointly influence an outcome in a way that cannot be fully explained by their lower-order effects, and are thus nonlinear functions of more than two variables. At the end of the day, what lies at the bottom of this debate is simply whether one aims to imputate as the source of the nonlinearity (e.g. the higher-order interaction) to an intrinsic nonlinear dynamics, or to a structural effect. 
One can then ask, what is more parsimonious as a modelling choice to characterise real-world complex dynamics: a ``simple'' interaction structure endowed with complicated nonlinear node (aka vertex) dynamics, or a more ``complicated'' interaction
structure supporting simpler (e.g.\ diffusive or linear) dynamics?

\medskip \noindent
To guide the intuition, consider the dynamical system defined on the nodes $i$ of a graph $G=(V,E)$:
\begin{equation}
\dot{x}_i = x_i + \prod_{j\in \partial_i} x_j,
    \label{eq:A}
\end{equation}
where $\partial_i$ is the topological neighborhood of node $i$. As a first disclaimer, note that while this is simply a nonlinear system of coupled ordinary differential equations, observe that calling this a graph-based dynamics is already a modelling choice, fueled by the convenient notational choice of using $\partial_i$. In other words, ontologically there is no graph, but we choose to write the system in a convenient way that implicitly allows for such an interpretation. Perhaps a more evident presence of a graph occurs where the whole graph structure --i.e. its adjacency matrix for labelled graphs-- is explicit in the formulation of the dynamics \cite{SIAM}, e.g.
\begin{equation}
\dot{x}_i = f(x_i) + \sum_{j=1}^{|V|} A_{ij} g(x_i,x_j),
    \label{eq:B}
\end{equation}
for some functions $f(\cdot),\ g(\cdot)$.
This is just a subset of all possible graph-based dynamics (albeit rather important for modelling real-world dynamics of different garments). By construction, it doesn't contain high-order interactions, so if we really want to explicitly incorporate the whole interaction structure in the dynamical equations, then in principle it seems legitimate to say that we need a higher-order network (in rigor, again Eq.~\ref{eq:B} per se does not really need that a graph exist, it is just convenient to interpret $A$ as the adjacency matrix of a graph).\\
Anyway, let's come back to Eq.~\ref{eq:A} and call this is a nonlinear graph-based dynamics, where the multilinear product $\prod_{j\in \partial_i} x_j$ is here the source of the nonlinearity. Observe that this is indeed a high-order interaction, in the sense that the whole topological neighborhood of node $i$ influences its evolution, and such influence cannot, in general, be decomposed into sums of two-body interactions. 
At the same time, Eq.~\ref{eq:A} can also be encoded within a hypergraph notation,  particularly as the high-order interaction $\prod_{j\in \partial_i} x_j$ can be seen as the state variable  of the hyperedge $\partial_i\equiv\{j\}_{j\in \partial_i}$ (using the definition of an hyperedge state as the product of states for the set of vertices forming such hyperedge). Such interpretation naturally requires augmenting the state space (which originally is formed by $x_i, \forall i=1,2,\dots,|V|)$ in a way that now such state space is spanned by the states of all hyperedges. Accordingly, Eq.~\ref{eq:A} allows for (at least) two equally valid yet different interpretations: one is a nonlinear evolution in a state space  $\mathbb{R}^{|V|}$, where $V$ is the vertex set of the interaction graph $G=(V,E)$ --a graph-based nonlinear dynamics--, whereas the other is an {\it apparently linear}\footnote{Later in the text we show that for this hypergraph-based linear evolution to be equivalent to the nonlinear graph-based one, some state variable closure conditions need to hold.} evolution in the state variables of the hypergraph induced by the full power set of the vertices of $G$.
There is no obvious way to decide a priori which interpretation is closer to reality.

\medskip \noindent
Authors in \cite{maxi} study scenarios where hypergraph-based dynamics can be reduced to graph-based ones, and indeed \cite{Peixoto} argues that such reducibility is generic. 
From our preceding example it seems intuitive to see that the apparent linear dynamics running on a hypergraph can be represented as a nonlinear (multilinear, hence with high-order interactions) dynamics running on a graph if we identify the hyperedge state variable with a nonlinear interaction between vertex state variables.
In passing, note that this ambiguity generates what almost seems like a social-epistemic conundrum: where do we ultimately locate the source complexity, in the dynamics or in the interaction topology?
Under representational equivalence, this would become a matter of explanatory commitment. Distinct scientific traditions privilege different primitives and therefore will probably exhibit systematic biases toward one representation over the other. For instance, dynamicists might be trained to think that ultimately dynamics determines behavior --there is neither a graph, nor a hypergraph, just a system of coupled nonlinear ODEs--; whereas network scientists might on the other hand be trained to think that structure shapes dynamics, and that topology (more is different) is ultimately the source of emergent complexity.
This simple observation should be taken into account when evaluating some of the otherwise valid critiques raised in \cite{Peixoto}.


\medskip Now, representation equivalence requires the two directions: while linear dynamics running on higher-order networks might be represented as graph-based nonlinear dynamics \cite{Peixoto, maxi}, the converse statement --i.e. what classes of nonlinear dynamics running on a graph can be represented as a {\it linear} dynamics on a richer higher-order network-- seems less evident, and it is the question we address here.
Our approach builds on classical results from dynamical systems theory, specially Carleman linearization \cite{Carleman1932}.  Some well-known representation theorems state that arbitrary complex ({\it nonlinear}) dynamics running on a finite-dimensional space can always be represented by a {\it linear} operator in a {\it suitably  augmented} (in general, Hilbert) space. When such space is constructed by adding sufficiently many tensor products of the original state space variables, such representation is guaranteed via Carleman embedding (aka Carleman linearization) theory \cite{Carleman1932}. A more general approach is followed for
the Koopman \cite{Koopman1931} or the Perron-Frobenius operators, that are applied, respectively, to generic functions of state variables --i.e. {\it observables}-- (Koopman) and densities (Perron-Frobenius), rather than on tensor products of state variables. Another conceptual similarity is that of linear chaos \cite{linearchaos}, i.e. some continuous {\it linear} operators on Hilbert space can be shown to be {\it chaotic} in the sense of Devaney, thereby breaking the traditional association between chaos and nonlinearity at the expense of working in infinite dimensions.\\
In all these cases, a similar Occam's razor-style question can be formulated as to what is a more parsimonious representation: a nonlinear evolution in low-dimensional state space, or a linear evolution in an augmented, possible infinite-dimensional space? While the question seems a priori to be a no-brainer --one shall use the representation that requires the smallest number of dimensions--, observe that parsimony not always boils down to minimizing the state-space dimension, and has to be traded-off with  algebraic simplicity or analytical tractability. For instance,
the existence of powerful linear-dynamical theories (spectral analysis, linear control, etc) is by no means on par with the arguably smaller range of tools at our disposal to characterize nonlinear dynamics. From another angle, each representation can give complementary insights. Think for instance of a simple one-dimensional nonlinear map: studying this directly as a nonlinear system in state space provides useful and direct insights on many orbit-related aspects such as bifurcations, attractors, Lyapunov exponents, and so on. The linear representation, on the other hand, involves considering the evolution of its density via the Perron-Frobenius (transfer) operator. This is a linear evolution (yet infinite-dimensional) and its analysis provides insights on statistical (ensemble) properties of the map, such as invariant densities, mixing rates, ergodicity, and the like, not directly accessible to an orbit-focused analysis. Both approaches have proven to be equally (and massively) successful in the field of nonlinear dynamics. All in all, probably the most useful representation is case-dependent.

\medskip
Let us resort now to Carleman-type arguments to address our problem, and assume that we define a high-order --or more generally, a nonlinear-- evolution on a finite  state space $x=(x_1,x_2,\dots x_{|V|})\in \mathbb{R}^{|V|}$. Carleman theory \cite{Carleman_theory} proceeds by lifting the state space $x \to (x,x^{\otimes 2},x^{\otimes 3},\dots)$, so that the augmented space is the direct sum of symmetric tensor powers of the original state space. For sufficiently large liftings (in general infinite-dimensional), the original nonlinear evolution becomes linear. Carleman approach is preferred here over e.g. a more general Koopman formalism \cite{Koopman_theory} as one can identify the monomials generated by Carleman lifting as the state variables of a higher-order network.\\
For illustration, start by considering a state space $x=(x_1,x_2,x_3)$, i.e. the state space of a dynamics running on a graph $G=(V,E)$ with $|V|=3$ vertices, where vertex $i$ has state variable $x_i$. Consider now the following nonlinear dynamics on $G$:
\begin{eqnarray}
&&\dot{x_1}=x_2\nonumber\\
&&\dot{x_2}=x_1 \label{eq:C} \\
&&\dot{x_3}=x_1x_2.\nonumber
\end{eqnarray}
This is explicitly a nonlinear graph-based dynamics with $E=\{(1,2),(2,1),(3,1),(3,2)\}$, i.e. the graph is directed. To linearise the dynamics, initially we need to define the lifted state-space as $(z_1,z_2,z_3,z_{12})$, where $z_i:=x_i, \ i=1,2,3$ and $z_{12}:=x_1x_2$ are, respectively, the state variables of the three vertices and the hyperedge $(1,2)$ (note that this last one is one of the terms in $x^{\otimes2}$). Then, we need to check whether the new state space is linearly closed under the time evolution. For continuous dynamics (ODEs), that means that the time derivative of any lifted state variable needs to be expressable as a linear combination of the rest of state variables. The state variables $z_i$ trivially fulfil this. Now $\dot{z_{12}}=\dot{x_1}x_2 + \dot{x_2}x_1 = x_2^2 + x_1^2$. Incidentally, note that neither $x_1^2$ nor $x_2^2$ can be identified with hyperedge states: we will come back to this problem later, and for argument sake, let us assume for now they can. Also, neither $x_1^2$ nor $x_2^2$  belong to the lifted state space, so we need to add them. Then, we further define $z_{11}:=x_1^2$ and $z_{22}:=x_2^2$, and add these to the lifted space (note that $x_1^2$ and $x_2^2$ are also terms from $x^{\otimes2}$). In this new lifted space, $\dot{z}_{12}= z_{11}+z_{22}$. Again, now we need to check linear closure for $z_{11}$ and $z_{22}$. We have $\dot{z}_{11}=2x_1\dot{x}_1=2x_1x_2=2z_{12}$, and likewise for $\dot{z}_{22}=2x_2\dot{x_2}=2x_2x_1=2z_{12}$. We now have linear closure for all state variables, and the process is finished. The linear dynamics, in matrix mode, reads
\begin{equation}
\dot{\left( \begin{array}{c} z_1 \\ z_2 \\ z_3 \\ z_{12} \\ z_{11} \\ z_{22} \end{array} \right)} = 
\left( \begin{array}{cccccc}
0 & 1 & 0 & 0 & 0 & 0 \\
1 & 0 & 0 & 0 & 0 & 0 \\
0 & 0 & 0 & 1 & 0 & 0 \\
0 & 0 & 0& 0 & 1 & 1 \\
0&0&0&2&0&0\\
0&0&0&2&0&0\\
\end{array} \right)
\left( \begin{array}{c} z_1 \\ z_2 \\ z_3 \\ z_{12} \\ z_{11} \\ z_{22} \end{array} \right) \label{eq:D}
\end{equation}
The example above illustrates the method, although note that the nonlinearity involved in Eq.~\ref{eq:C} is not higher-order.
Let us now illustrate the process of linearizing Eq.~\ref{eq:A} --where higher-order interactions are explicit-- in a very simple (directed) graph topology of $|V|=4$ vertices where $\partial_4 = (1,2,3)$ and $\partial_1 = \partial_2 = \partial_3 = \emptyset$, i.e
\begin{eqnarray}
&&\dot{x_1}=x_1\nonumber\\
&&\dot{x_2}=x_2 \label{eq:E} \\
&&\dot{x_3}=x_3\nonumber\\
&&\dot{x_4} = x_4 + x_1x_2x_3.\nonumber
\end{eqnarray}
It is easy to see that this dynamics is linearly closed in the vertex state variables $z_i=x_i$ for $i=1,2,3,4$ plus the lift $z_{123}=x_1x_2x_3$ which is identified with the state variable of the hyperedge $(1,2,3)$, since this extra variable is linearly closed under time evolution $\dot{z}_{123}=3z_{123}$. In matrix form, the hypergraph-based linear dynamics representation is:
\begin{equation}
\dot{\left( \begin{array}{c} z_1 \\ z_2 \\ z_3 \\ z_{4} \\ z_{123}  \end{array} \right)} = 
\left( \begin{array}{ccccc}
1 & 0 & 0 & 0 & 0  \\
0 & 1 & 0 & 0 & 0  \\
0 & 0 & 1 & 0 & 0  \\
0 & 0 & 0& 1 & 1  \\
0&0&0&0&3\\
\end{array} \right)
\left( \begin{array}{c} z_1 \\ z_2 \\ z_3 \\ z_{4} \\ z_{123} \\ \end{array} \right) \label{eq:F}
\end{equation}

\medskip \noindent {\bf From hypergraphs to hb-graphs --} All the possible hyperedge state variables of order up to $|V|$ in a hypergraph defined on $|V|$ vertices can be found in the components of the liftings up to $x^{\otimes |V|}$, but the example in Eq.~\ref{eq:C} made it clear that not all the terms arising in Carleman lifting can indeed be identified with hyperedge states, only multilinear forms $\prod_{j}x_j$ can. 
It is already in this sense that one can claim that {\it only some} graph-based dynamics with high-order interactions can be represented as a linear dynamics in a {\it standard} hypergraph (such as Eq.~\ref{eq:E}). Consider for instance the term $x_1^2$, that appears in the linearisation of Eq.~\ref{eq:C} as one of the terms in $x^{\otimes 2}$. $x_1^2$ is not the state variable of a 1-hyperedge, but can be interpreted as a ``self-interecting'' 2-hyperedge $(1,1)$, which is a degenerate notion of a 2-hyperedge. Likewise, the term $x_1^2x_2$, emerging from the coordinates of $x^{\otimes 3}$, is not the state variable of any 3-hyperedge\footnote{Of course, it can also be interpreted as the product of the states of hyperedges $(1,2)$ and $(1)$, but we don't consider this possibility as we are aiming to build a linear representation.}, but can be interpreted as a degenerate notion of a 3-hyperedge if we allow for vertex multiplicities. While standard hyperedges do not allow for vertex multiplicities,  there exist a more flexible combinatorial structure, called hyper-bag-graphs (hb-graphs) \cite{hb, hb2}, that precisely allow hyperedges to have such multiset indices. hb-graphs enhance the properties of hypergraphs with a multigraph notion: just like hypergraphs generalise graphs in the sense that the hyperedges are sets of multiple --yet distinct-- nodes and multigraphs generalise graphs in the sense that pairs of vertices can be connected by multiple edges, hb-graphs merge these two generalisations and allow for hyperedges to have repeated indices. Observe that hb-graphs are seldom used within the network science community, with some notable exceptions \cite{Bick}.\\
Within this more general combinatorial structure, it is then easy to see that any state variable appearing in an arbitrary lifting $x^{\otimes k}$ is naturally associated to the state variable of a particular hb-edge: for instance, the term $x_1^2x_2$ is the state variable of the hb-edge $(1,1,2)$.
Accordingly, we now have a strategy to identify any possible term arising in the Carleman linearization of a nonlinear graph-based dynamics with the state variable of an hb-edge. There is just one missing aspect to consider: what is the number of necessary hb-edges to exactly represent a nonlinear graph-based dynamics as a linear hb-graph-based one?

\medskip \noindent {\bf All you need is closure --} As already advanced, in order for the linearisation to represent exactly the original nonlinear graph-based dynamics, the lifted state space must itself be linearly closed under time evolution. Closing this space often requires to iteratively (i) include in the lifted space new elements from $x^{\otimes k}$, (ii) check if these new state variables are linearly closed under time evolution, and (iii) add more state variables until the system is closed. The linearisation of Eq.~\ref{eq:C} employed such procedure. Now, if the process is fulfilled after a finite number of steps, then the hb-graph is of finite size. However, if the system never closes, Carleman linearisation is exact only in the limit of infinite-dimensional lifting: in this case the resulting hb-graph, while being of order $|V|$ (i.e. having the same number of vertices as the graph $G$), is of infinite size (infinite number of hb-edges). This situation emerges already in Eq.~\ref{eq:A} for graph topologies slightly more complex than the one used in Eq.~\ref{eq:E}, for instance if $G$ is a complete graph, every new hb-edge state included in the lift requires
 the inclusion of more hb-edge states to guarantee closure. This hierarchy never ends, hence closure is only guaranteed asymptotically.
Observe that, in contrast to hypergraphs defined on $|V|$ vertices, where the number of hyperedges is bounded by $2^{|V|}-1$, hb-graphs are able to include an unbounded amount of hb-edges precisely thanks to its multiset structure. 
Of course, in the event one aims to keep the number of hb-edges finite, then in general the linearization of the dynamics on the finite-size hb-graph will only be approximate (effectively a truncation). Accurate approximations can be obtained by taking so-called finite-sections \cite{amini}.

\medskip \noindent
As a last comment, observe that the methodology described so far apply for multilinear or finite polynomial nonlinearities defined in the vertex set of the graph. For even more generic (analytic) nonlinear dynamics defined on a graph with $|V|$ vertices, it is easy to see that there also exists a hb-graph with the same $|V|$ vertices that exactly linearise the dynamics, since an analytic nonlinearity admits an (infinite) power series expansion. Of course, in this case the number of hb-edges needed (i.e. the size of the hb-graph) is always infinite.

\medskip
To summarise, we leverage Carleman linearisation theory to explicitly construct a correspondence between dynamics and structure in network dynamics, by which nonlinear dynamics (of which high-order interactions are a subset) defined purely on the vertex set of a graph can be represented as a linear dynamics on a sufficiently richer combinatorial structure. 
When the dynamics defined in the vertices $V$ of the graph is multilinear, its representation in terms of a linear dynamics in a hypergraph of order $V$ is possible as long as only multilinear terms are needed in the lifted space (and closure holds), otherwise the linear representation requires a generalization of the hypergraph known as a hb-graph. The total number of hb-edges (i.e. the size of the hb-graph) needed for the representation to be exact depends on the concrete nonlinearity: in some cases this number is finite and small (as in Eqs.~\ref{eq:C},\ref{eq:E}), but for generic graph topologies with multilinear, polynomial nonlinearities or more complex analytic ones, an infinite number of hb-edges is needed, or otherwise the finite truncation provides only an approximate representation. It is difficult, in practice, to tell which specific nonlinear dynamics can be linearised in finite space, but what seems to be clear is that virtually all {\it interesting} nonlinearities (i.e. nontrivial attractors such as multiple fixed points, chaotic attractors, etc) will need infinite-dimensional liftings --and thus infinite-size hb-graph linearisation--, if only because linear systems in {\it finite} dimensions simply do not show these type of attractors, so in general a topological conjugacy that maps the two representations won't exist. 

\medskip \noindent Three final remarks are in order. First, one might ask why not to perform Carleman linearisation directly in graph space, without the needs to resort to more complex combinatorial structures. The answer is simple: because the lifted state variables required for the linearisation do not naturally relate to vertex state variables, but find a natural interpretation as hyperedges or hb-edges. Second, our analysis focuses on graph-based continuous-time dynamics (ODEs), although similar results hold for the discrete-time case (coupled maps), where in the latter case linear closure of state variables under time evolution needs to be assessed in terms of map composition, rather than time derivative. Third, observe that in this work we focus on the linearisation problem of nonlinear graph-based dynamics. If we instead asked about whether nonlinear graph-based dynamics can be represented by a {\it nonlinear} evolution in a higher-order network, then it is easy to see that this is true: a simple strategy is to adequately reinterpret any nonlinear function $f(x_i, \partial_i)$ (acting on a arbitrary vertex $i$ of a graph with $|V|$ vertices and its topological neighborhood) as the same nonlinear function acting on a suitable combination of hyperedges of a hypergraph defined in the same vertex set, i.e. the concrete set of hyperedges are trivially induced by the form of $f$ and $G$, as pointed in \cite{Peixoto}.\\
All in all, the existence of the representation duality discussed here provides a nuanced view on the graph vs high-order network debate \cite{Peixoto}. Just like investigating chaotic dynamics either directly in (finite-dimensional) state space or in (infinite-dimensional) distribution space via Perron-Frobenius both have their merits, studying nonlinear (higher-order) dynamics on graphs vs their linear version running on a higher-order network structure could potentially yield complementary insights. Among other possibilities, the linear representation can be leveraged to design strategies for the nonlinear optimal control \cite{control} of nonlinear graph-based dynamics \cite{control2}.


\medskip
\noindent {\bf Acknowledgments --} I wish to thank feedback from many colleagues in the network science community on a preliminary version of this work, including M. San Miguel for pointing me to \cite{maxi}, C. Bick for pointing me to coupled cell hypernetworks \cite{Bick}, and T. Gross for a question on Carleman closure.
The author acknowledges partial support from project CSxAI (PID2024-157526NB-I00) funded by MICIU/AEI/10.13039/501100011033/FEDER, UE, from a Maria de Maeztu project (CEX2021-001164-M) funded by the MICIU/AEI/10.13039/501100011033, and from the European Commission Chips Joint Undertaking project No. 101194363 (NEHIL).\\

\bibliographystyle{unsrt}

\begin{thebibliography}{10}

\bibitem{Battiston2020}
F. Battiston, G. Cencetti, I. Iacopini, V. Latora, M. Lucas, A. Patania, J-G Young, and G. Petri. 
Networks beyond pairwise interactions:
Structure and dynamics,
{\it Physics Reports} 874 (2020), 1--92.

\bibitem{SIAM} C. Bick, E. Gross, H.A. Harrington, H. A., and M.T. Schaub. What are higher-order networks?. {\it SIAM review} 65, 3 (2023): 686-731.

\bibitem{Battiston25}
F. Battiston, E. Amico, A. Barrat, G. Bianconi, G. Ferraz de Arruda, B. Franceschiello, I. Iacopini, S. Kéfi, V. Latora, Y. Moreno, M.M. Murray, T.P. Peixoto, F. Vaccarino and G. Petri, 
The physics of higher-order interactions in complex systems. {\it Nature physics} 17, 10 (2021).

\bibitem{maxi} J. Llabrés, R. Toral, M. San Miguel, and F. Vázquez (2026). Reducibility of higher-order to pairwise interactions: Social impact models on hypergraphs. arXiv preprint arXiv:2601.05169.

\bibitem{lambiotte} L. Neuhäuser, A. Mellor, A. and R. Lambiotte. Multibody interactions and nonlinear consensus dynamics on networked systems. {\it Physical Review E} 101, 32 (2020), 032310.



\bibitem{Peixoto} T.P. Peixoto, L. Peel, T. Gross, M. de Domenico, Graphs are maximally expressive for higher-order interactions, arXiv:2602.16937v1



\bibitem{Carleman1932}
T. Carleman, Application de la théorie des équations intégrales linéaires
aux systèmes d'équations différentielles non linéaires,
{\it Acta Mathematica} 59 (1932), 63--87.

\bibitem{Koopman1931}
B. O. Koopman,
Hamiltonian systems and transformation in Hilbert space,
{\it Proceedings of the National Academy of Sciences of the USA} 17 (1931), 315--318.

\bibitem{Carleman_theory} K. Kowalski, and W.H. Steeb (Eds.) {\it Nonlinear dynamical systems and Carleman linearization} (World Scientific, 1991).

\bibitem{Koopman_theory} 
S.L. Brunton, M. Marko Budišić, E. Kaiser, and J. N. Kutz. Modern Koopman Theory for Dynamical Systems. {\it SIAM Review} 64, 2 (2022)

\bibitem{carleman2} M. Forets, A. Pouly. (2017). Explicit error bounds for Carleman linearization. arXiv preprint arXiv:1711.02552.



\bibitem{linearchaos} K.G. Grosse-Erdmann, A.P. Manguillot. {\it Linear chaos} (Springer Science \& Business Media, 2011).

\bibitem{hb} X. Ouvrard, J.M.L Goff, and S.  Marchand-Maillet, S (2018). Adjacency and tensor representation in general hypergraphs. part 2: Multisets, hb-graphs and related e-adjacency tensors. arXiv preprint arXiv:1805.11952.

\bibitem{hb2} X. Ouvrard, J.M. Le Goff, and S. Marchand-Maillet. Exchange-based diffusion in Hb-Graphs: Highlighting complex relationships in multimedia collections (extended version). {\it Multimedia Tools and Applications} 80(15) (2021): 22429-22464.

\bibitem{Bick} M. Aguiar, C. Bick, C., and A. Dias. Network dynamics with higher-order interactions: coupled cell hypernetworks for identical cells and synchrony. {\it Nonlinearity} 36, 9 (2023): 4641-4673.

\bibitem{amini} A. Amini, C. Zheng, Q. Sun, and N. Motee. Carleman linearization of nonlinear systems and its finite-section approximations. arXiv preprint arXiv:2207.07755.

\bibitem{control}S.L. Brunton, B.W. Brunton, J.L. Proctor, and J.N Kutz. Koopman invariant subspaces and finite linear representations of nonlinear dynamical systems for control. {\it PloS one}, 11(2) (2016): e0150171.

\bibitem{control2} S.P. Cornelius, W.L. Kath, and A.E. Motter. Realistic control of network dynamics. {\it Nature communications} 4, 1 (2013).

\end{thebibliography}

\end{document}